\documentclass{article}

\usepackage{arxiv}

\usepackage[utf8]{inputenc} % allow utf-8 input
\usepackage[T1]{fontenc}    % use 8-bit T1 fonts
\usepackage{hyperref}       % hyperlinks
\usepackage{url}            % simple URL typesetting
\usepackage{booktabs}       % professional-quality tables
\usepackage{amsfonts}       % blackboard math symbols
\usepackage{nicefrac}       % compact symbols for 1/2, etc.
\usepackage{microtype}      % microtypography
\usepackage{lipsum}
\usepackage{graphicx}
\usepackage{subcaption}
\graphicspath{ {./Figures/} }

\title{NASA-TLX Web App: An Online Tool to Analyse Subjective Workload}

\author{
 Vinoth Pandian Sermuga Pandian \\
  Fraunhofer FIT\\
  Sankt Augustin\\
  Germany\\
  \texttt{pandian@fit.fraunhofer.de} \\
   \And
 Sarah Suleri \\
  Fraunhofer FIT\\
  Sankt Augustin\\
  Germany\\
  \texttt{sarah.suleri@rwth-aachen.de} \\
}

\begin{document}
\maketitle
\begin{abstract}
NASA Task Load Index (NASA-TLX) is a widely used assessment technique to compute subjective workload experienced during a task. It evaluates workload using six dimensions: mental demand, physical demand, temporal demand, frustration, effort, and performance. This paper presents a web app to assist experimenters in using NASA-TLX to commute subjective workload. The web app enables the experimenter to conduct various experiments simultaneously and offers the participants a concise interface to provide their subjective evaluation. It performs the calculations at the backend and provides the computed results comprehensively. The web app provides a dashboard for the experimenter to visualize and export the summary of results. Qualitative feedback from 12 experimenters indicated that the NASA-TLX web app is relevant, helpful, and easy to use.
\end{abstract}

% keywords can be removed

\section{Introduction}
NASA-TLX is a widely used multidimensional assessment tool that evaluates the perceived workload for any given task \cite{NASA2017}. The task is assessed on the basis of the participant's subjective rating (0 to 100) of six dimensions: mental demand, physical demand, temporal demand, performance, effort, and frustration.

As a first step, participants provide their subjective ratings for each dimension. Then, they are asked to assign weights to these dimensions in a pairwise comparison. The purpose of this step is to evaluate the perceived importance of each dimension. The assigned weights are then multiplied with the subjective rating given by the participant for the respective dimension, thus computing the cumulative subjective workload for any given task.

\section{Literature Review}
By a keyword search for existing software and websites to conduct a computerized NASA-TLX study, we identified nine relevant tools. Excluding the tools that required paid-license \cite{millisecond} or source code compilation before usage \cite{jnasatlx, jmpolomN55:online}; we reviewed six remaining tools: four websites and two applications.
    
Specifically, two websites \cite{keithv, asavva} allow conducting NASA-TLX analysis on a single participant and display the individual result. Whereas, the other two \cite{isellsoa14:online, Sharek2011} support interviewing multiple participants for an experiment and provide an overview of their results. Only one website \cite{Sharek2011} synchronizes experiment results in an online database and supports experiment data export.

Two applications support workload analysis offline: NASA-TLX official app by NASA's Ames Research Center \cite{NASA_TLX}; and NASA-TLX desktop app by Playgraph \cite{nasatlx-desktop}. The official app supports iOS mobile devices \cite{NASA_TLX}. Whereas, Playgraph's NASA-TLX app supports all desktop platforms that can execute Adobe Air runtime \cite{nasatlx-desktop}. Both apps support workload analysis of multiple participants for an experiment, online data synchronization, and experiment data export. Also, the official NASA-TLX app provides a graph of each participant's data \cite{NASA_TLX}.

While these tools aid to computerize paper and pencil version of NASA-TLX, we believe it would be beneficial to provide an overview of experiment data by visualizing it for quick inference of task load trends.

\begin{figure}[t]
    \centering
     \begin{subfigure}[c]{0.45\textwidth}
        \includegraphics[width=\textwidth]{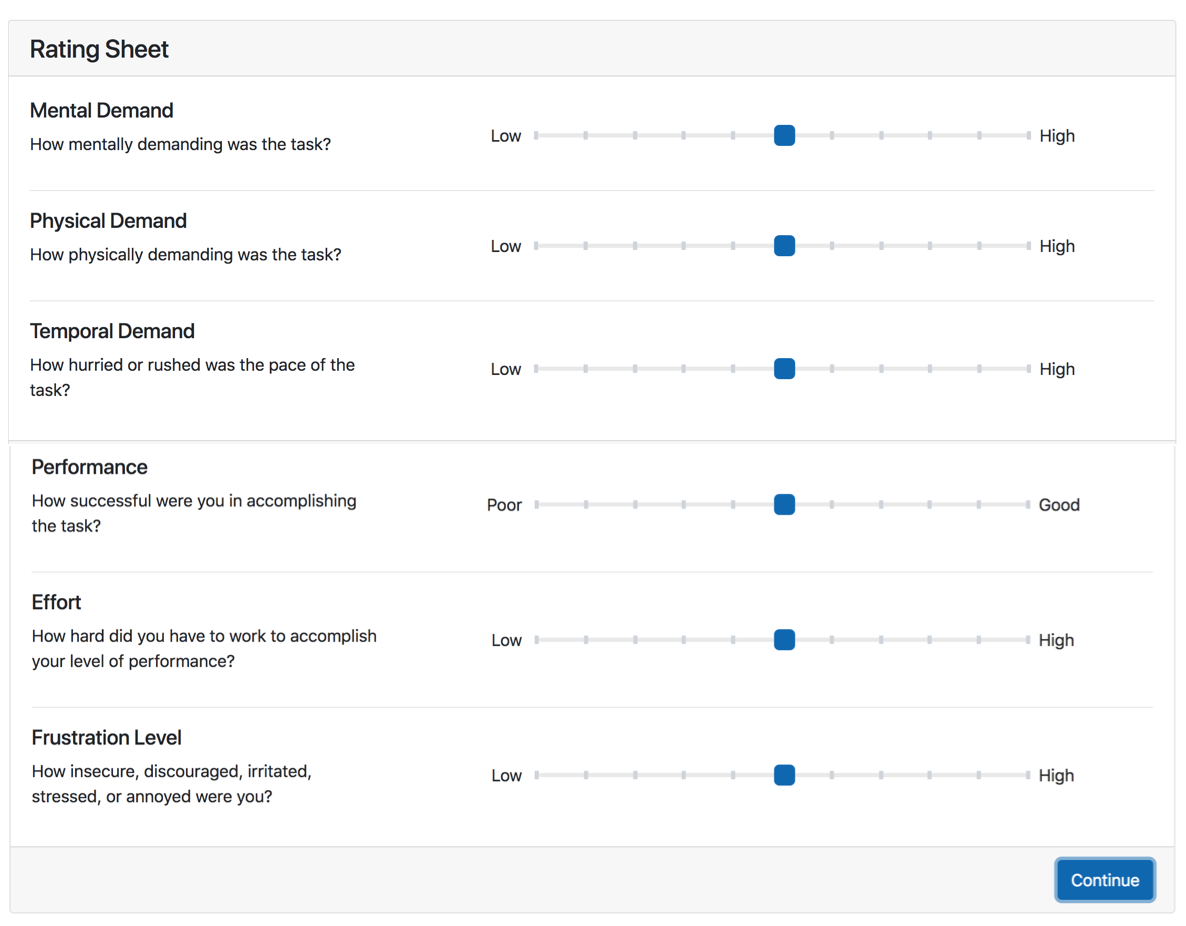}
        \caption{Rating sheet for various dimensions of NASA-TLX}
        \label{fig:rating}
     \end{subfigure}
     \begin{subfigure}[c]{0.45\textwidth}
        \includegraphics[width=\textwidth]{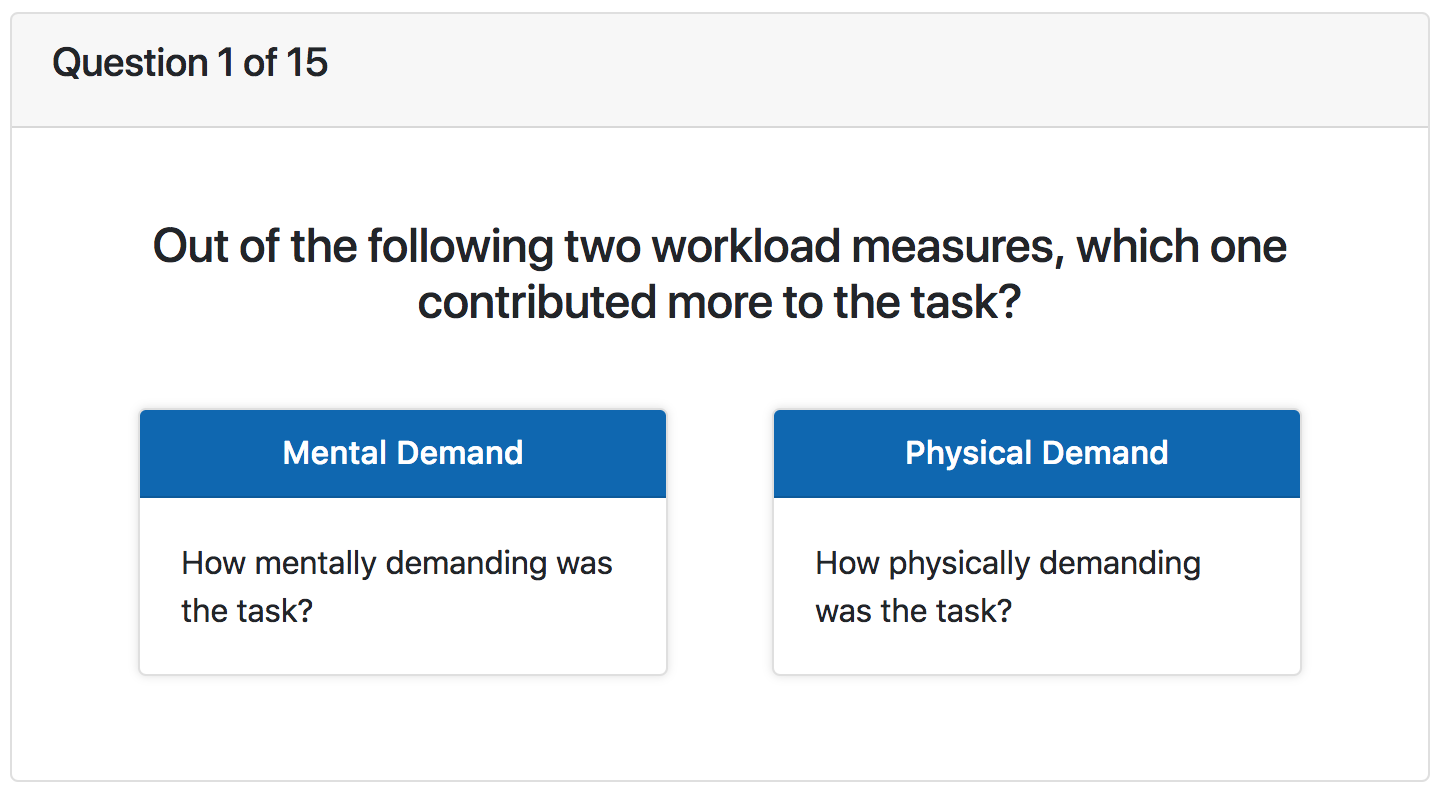}
        \setlength{\abovecaptionskip}{40pt plus 5pt minus 0pt}
        \caption{Workload dimension comparison card}
        \label{fig:q}
     \end{subfigure}
     
        \caption{Rating sheet and workload dimension comparison forms of NASA-TLX web app}
        \label{fig:three graphs}
\end{figure}

\section{NASA-TLX Web Application}
We present a web application\footnote{\url{https://nasa-tlx.firebaseapp.com/}} \footnote{\url{https://github.com/vinothpandian/nasa-tlx-web}} to perform subjective workload analysis using NASA-TLX. This application aims to assist experimenters in conducting the subject workload study and analyze with ease. The web app manages multiple experiments simultaneously with respect to experiment ID and assigns each participant with a randomly generated unique id. 

\begin{figure}[t]
\centering
\includegraphics[width= \columnwidth]{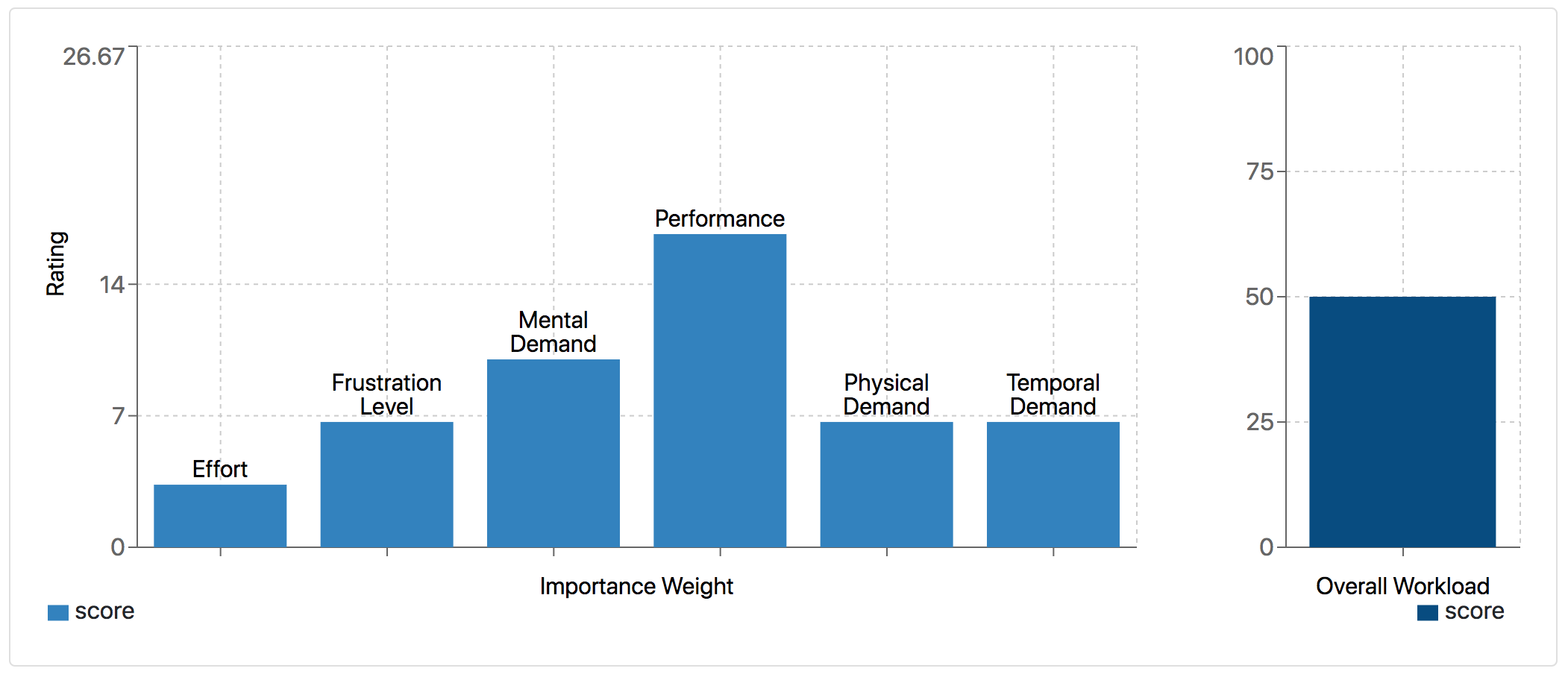}
\caption{Summary of automatically computed results in a graphical format.}
\label{fig:result}
\end{figure}

Participants are provided with a brief description of each dimension of NASA-TLX and then asked to provide the respective ratings using a slider (Figure \ref{fig:rating}). In the next step, participants assign comparative weight to each dimension (Figure \ref{fig:q}). Once these subjective values are collected, the respective results of subjective workload analysis are automatically generated (Figure \ref{fig:result}). The experimenter can access the results as raw data and graphical summary for each participant as well as for the entire experiment. These results can also be exported as CSV or JSON file.

\section{Implementation}
We developed NASA-TLX as an open-source web app in order to make it accessible via various platforms. It supports almost all of the major browsers: Mozilla Firefox, Google Chrome, Safari, and Opera, except Microsoft Edge and Internet Explorer.

\subsection{User Interface Design}
We designed the User Interface (UI) with Twitter's Bootstrap components \cite{Bootstra76:online}. Bootstrap is an open-source toolkit for building responsive, mobile-first web applications. With Bootstrap at its core, the NASA-TLX web app adapts seamlessly to any device resolution and orientation. Although it is designed to work well in portrait mode, we recommend to use it in landscape orientation as it provides a much larger screen area for participants to interact with the sliders in the rating sheet (Figure \ref{fig:rating}). 

\subsection{Front-end}
The front-end of NASA-TLX web app is written using JavaScript (JS). The UI components are built using Facebook's React library. React is an open-source component-based library to build interactive UI \cite{ReactAJa46:online}. In addition, we used few popular JS packages for web app development: \textit{React-Router} for managing the page navigation \cite{ReactRou38:online}; \textit{Redux} for maintaining page state \cite{ReadMeRe22:online}; and \textit{Redux-Saga} for managing asynchronous data fetching. 

\subsection{Back-end}
The back-end of the NASA-TLX web app uses Google's Firebase Realtime Database. It is a No-SQL cloud-hosted database that allows storing and syncing data in real-time across all devices and platforms \cite{Firebase60:online}. For privacy and security, the web app uses a Firebase authentication service. It securely stores experiment data in the cloud and restricts access for everyone except the authenticated user. Firebase authentication supports login using existing Google, Facebook, or Twitter accounts; users can also register using email \cite{Firebase38:online}.  

\subsection{Optimization \& Performance}
NASA-TLX web app is developed using a webpack module bundler.  Webpack bundles the JS code into fewer files and optimizes them to improve the web app's loading time \cite{webpack40:online}. With the webpack, the web app's source code is optimized, minified, and compressed to load faster. 

The web app is deployed in Google's Firebase hosting service. It is a fast and secure web hosting provider and content delivery network \cite{Firebase33:online}. Each script is set to load asynchronously whenever required; as a result, the web app loads quickly and performs smoothly even in a slow internet connection. In totality, the complete site usage requires 701KB internet data in browsers with gzip compressions support. Also, it leverages browser-caching, so once the site is loaded, it performs fast, until its cache is removed. 

\section{Qualitative Feedback}
We collected qualitative feedback from 12 experimenters who had previously used NASA-TLX paper questionnaires in multiple experiments. The experimenter was asked to use the NASA-TLX web app instead of a paper questionnaire and conduct a workload experiment. The feedback indicated that experimenters found the NASA-TLX web app to be relevant, convenient, and easy to use.

\section{Conclusion}
We presented a web app to assist experimenters in conducting workload analysis using NASA-TLX. The web app provides the participants with an interface to submit their subjective evaluation. The web app computes the results automatically based on the scale values and weights submitted by the participants. It also provides a dashboard for the experimenter to view and export the summary of results for various experiments. A preliminary study showed that experimenters found this app convenient and easy to use. In the future, we intend to extend this research by including offline data persistence and multiple language support.

%ACKNOWLEDGMENTS 
\section{Acknowledgments}
We would like to extend our gratitude to Nilda Kipi and Svetlana Shishkovets for their help and support.

\bibliographystyle{unsrt}  
\bibliography{references}  %%% Remove comment to use the external .bib file (using bibtex).

\begin{thebibliography}{10}

\bibitem{NASA2017}
Sandra~G Hart and Lowell~E Staveland.
\newblock Development of nasa-tlx (task load index): Results of empirical and
  theoretical research.
\newblock In {\em Advances in psychology}, volume~52, pages 139--183. Elsevier,
  1988.
\newblock
  \url{https://www.sciencedirect.com/science/article/pii/S0166411508623869}.

\bibitem{millisecond}
{NASA Task Load Index (NASA-TLX)}.
\newblock \url{https://www.millisecond.com/download/library/nasatlx/}, 2010.

\bibitem{jnasatlx}
{Nasa TLX Test in Java Swing}.
\newblock \url{https://github.com/javanasatlx/jnasatlx/}, 2015.

\bibitem{jmpolomN55:online}
{NASA-TLX: Calculate the NASA Task Load Index (TLX) with this handy wxPython
  GUI}.
\newblock \url{https://github.com/jmpolom/NASA-TLX}, 2011.

\bibitem{keithv}
Keith Vertanen.
\newblock {NASA-TLX in HTML and JavaScript}.
\newblock \url{https://www.keithv.com/software/nasatlx/}.

\bibitem{asavva}
{NASA TLX - Online Software}.
\newblock \url{http://nasa.asavva.net/}.

\bibitem{isellsoa14:online}
{An implementation of the NASA Task Load Index (NASA-TLX)}.
\newblock \url{https://github.com/isellsoap/nasa-tlx}, 2012.

\bibitem{Sharek2011}
David Sharek.
\newblock {A useable, online NASA-TLX tool}.
\newblock In {\em Proceedings of the Human Factors and Ergonomics Society},
  pages 1375--1379, 2011.
\newblock \url{http://www.nasatlx.com}.

\bibitem{NASA_TLX}
{NASA TLX official App - NASA's Ames Research Center}.
\newblock \url{https://humansystems.arc.nasa.gov/groups/TLX/tlxapp.php}.

\bibitem{nasatlx-desktop}
{NASA-TLX Desktop app}.
\newblock \url{http://www.playgraph.com/apps.updates/NASA-TLX.air}, 2009.

\bibitem{Bootstra76:online}
Bootstrap - the most popular html, css, and js library in the world.
\newblock \url{https://getbootstrap.com/}.
\newblock (Accessed on 27/01/2020).

\bibitem{ReactAJa46:online}
React - a javascript library for building user interfaces.
\newblock \url{https://reactjs.org/}.
\newblock (Accessed on 27/01/2020).

\bibitem{ReactRou38:online}
React router: Declarative routing for react.js.
\newblock \url{https://reacttraining.com/react-router/web/guides/philosophy}.
\newblock (Accessed on 27/01/2020).

\bibitem{ReadMeRe22:online}
Redux - a predictable state container for javascript apps.
\newblock \url{https://redux.js.org/}.
\newblock (Accessed on 27/01/2020).

\bibitem{Firebase60:online}
Firebase realtime database | firebase realtime database | firebase.
\newblock \url{https://firebase.google.com/docs/database/}.
\newblock (Accessed on 27/01/2020).

\bibitem{Firebase38:online}
Firebase authentication | firebase.
\newblock \url{https://firebase.google.com/docs/auth/}.
\newblock (Accessed on 27/01/2020).

\bibitem{webpack40:online}
Webpack - a static module bundler for modern javascript applications.
\newblock \url{https://webpack.js.org/}.
\newblock (Accessed on 27/01/2020).

\bibitem{Firebase33:online}
Firebase hosting | firebase.
\newblock \url{https://firebase.google.com/docs/hosting/}.
\newblock (Accessed on 07/06/2018).

\end{thebibliography}

\end{document}